\documentstyle[sprocl]{article}

\def\bea{\begin{eqnarray}}
\def\eea{\end{eqnarray}}

\def\beq{\begin{equation}}
\def\eeq{\end{equation}}

\def\ga{\gamma}
\def\de{\delta}

\def\la{\lambda}

\def\si{\sigma}

\def\ps{\psi}
\def\om{\omega}
\def\Ga{\Gamma}

\def\mn{{\mu\nu}}

\def\half{{\textstyle{1\over 2}}}
\def\frac#1#2{{\textstyle{{#1}\over {#2}}}}

\def\lsim{\mathrel{\rlap{\lower4pt\hbox{\hskip1pt$\sim$}}
    \raise1pt\hbox{$<$}}}
\def\gsim{\mathrel{\rlap{\lower4pt\hbox{\hskip1pt$\sim$}}
    \raise1pt\hbox{$>$}}}
\def\sqr#1#2{{\vcenter{\vbox{\hrule height.#2pt
         \hbox{\vrule width.#2pt height#1pt \kern#1pt
         \vrule width.#2pt}
         \hrule height.#2pt}}}}

\begin{document}

\title{PROBING THE PLANCK SCALE IN \\
LOW-ENERGY ATOMIC PHYSICS }

\author{R.\ BLUHM}

\address{Physics Department, Colby College,
Waterville, ME 04901, USA\\E-mail: rtbluhm@colby.edu}


\maketitle\abstracts{
Experiments in atomic physics have exceptional sensitivity
to small shifts in energy in an atom, ion, or bound particle.
They are particularly well suited to search for unique low-energy
signatures of new physics,
including effects that could originate from the Planck scale.
A number of recent experiments have used CPT and Lorentz violation as a
candidate signal of new physics originating from the Planck scale.
A discussion of these experiments and their theoretical implications is
presented.
}

\section{Introduction}

It is known that our current understanding of particle interactions
as described by the standard model must break down at the Planck scale.
This is because the standard model ignores the effects of gravity,
which necessarily come into play at the Planck scale
$M_{\rm Pl} = \sqrt{\hbar c/G} \simeq 10^{19}$ GeV.
Much of the current work in theoretical high-energy physics is
devoted to finding a new fundamental theory that describes
physical interactions at the Planck scale.
Promising insights are being found in the context of string theory,
D-branes, and theories of quantum gravity.
Many of these theories include effects that violate
assumptions of the standard model,
including higher dimensions of spacetime,
unusual geometries, nonpointlike interactions,
and new forms of symmetry breaking.

A common misconception is that since the Planck scale is so
much higher than current accelerator energies,
physics at the Planck scale is inaccessible in experiments.
However,
this view is shortsighted because it fails to take into account
that experiments can be performed at exceptionally low energy
and are therefore potentially sensitive to effects from the
Planck scale that are heavily suppressed at ordinary energies.
For example, experiments in atomic physics are routinely sensitive
to small frequency shifts at the level of 1 mHz or less.
If this is due to an energy shift expressed in GeV,
it corresponds to a sensitivity of approximately
$4 \times 10^{-27}$ GeV.
Such a sensitivity is well within the range of energy one
might associate with suppression factors originating
>from the Planck scale.
For example,
the fraction $m_p/M_{\rm Pl}$ multiplying the proton mass
yields an energy of approximately $10^{-19}$ GeV,
while for the electron the fraction $m_e/M_{\rm Pl}$ times
the electron mass is about $2.5 \times 10^{-26}$ GeV.

Atomic physics has a rich history of testing for low-energy
signals of effects originating from high energy.
Examples include high-precision tests of quantum electrodynamics
as well as current efforts to measure atomic parity violation
and electric dipole moments in atoms.
The latter effects are expected to occur in the context of the standard
model and are associated with suppression factors
originating from the electroweak scale.
In order for a low-energy signal originating from the Planck scale
to be detectable in an atomic experiment and not be drowned out
by a less suppressed (permissible) signal,
it would necessarily have to involve corrections that cannot
be mimicked in the context of the standard model.

A promising set of candidate signals that could provide unambiguous
evidence of new physics originating from the Planck scale is
CPT and Lorentz violation.
These violations are forbidden in the standard model.
However,
it has been shown in the context of string theory
that violations of these symmetries can occur.
Based on these ideas,
a number of recent experiments in atomic physics have searched for
CPT and Lorentz violation as a signal of new physics originating
>from the Planck scale.
Many have obtained bounds that are well within the range
associated with suppressions from the Planck scale.

In the following sections,
I will review the theoretical ideas that motivate these
searches for CPT and Lorentz violation.
I will then briefly discuss a number of the atomic
experiments that have been performed in the last two to three years.
Additional details can be found as well in many of the other
articles in this volume.

\section{CPT and Lorentz Symmetry}

In the context of the standard model,
Lorentz symmetry and CPT are exact fundamental symmetries of nature.
In addition,
these symmetries are linked by the CPT theorem,
which states that all local relativistic field theories of
point particles are symmetric under CPT.
A prediction of the CPT theorem is that particles and antiparticles should
have exactly equal lifetimes, masses, and magnetic moments.

It has,
however,
been known for well over a decade that string theory can
lead to violations of CPT and Lorentz symmetry.\cite{kskp}
This is because strings are nonpointlike and have nonlocal interactions.
They can therefore evade the CPT theorem.
There are also mechanisms in string theory that can induce
spontaneous breaking of CPT and Lorentz symmetry.
This is due to certain types of interactions in string theory
among Lorentz-tensor fields that can destabilize the naive vacuum
and generate nonzero vacuum expectation values for Lorentz tensors.
The vacuum expectation values fill the true vacuum and cause spontaneous
Lorentz breaking.
This mechanism also induces spontaneous CPT violation whenever the
tensor-field expectation values involve an odd number of spacetime indices.
It has also been shown that geometries with noncommutative
coordinates can arise naturally in string theory\cite{connes98}
and that Lorentz violation is intrinsic to noncommutative field
theories.\cite{chlkO01}

A useful theoretical tool for studying CPT and Lorentz violation
is the standard-model extension.\cite{ck,kl01}
It provides a consistent theoretical framework that includes the
standard model (and SU(3)$\times$SU(2)$\times$U(1) gauge invariance)
and which allows for small violations of Lorentz and CPT symmetry.
It has been shown that any realistic noncommutative field theory
is equivalent to a subset of the standard-model extension.\cite{chlkO01}
To consider experiments in atomic physics it suffices
to restrict the standard-model extension to its QED sector
and to include only terms that are power-counting renormalizable.
The resulting QED extension has energy-momentum conservation,
the usual spin-statistics connection, and observer Lorentz covariance.
The renormalizability of the QED extension has recently
been shown to hold to one-loop.\cite{klp01}
The theory has also been used to study scattering cross
sections of electrons and positrons in the presence of
CPT and Lorentz violation.\cite{ck01}

The modified Dirac equation in the QED extension describing a
four-component spinor field $\ps$
of mass $m$
and charge $q = -|e|$ in an electric potential $A^\mu$ is
\beq
( i \Ga^\mu D_\mu - M) \ps = 0
\quad ,
\label{dirac}
\eeq
where
\beq
\Ga_\nu = \ga_\nu + c_\mn \ga^\mu + d_\mn \ga_5 \ga^\mu
\quad .
\label{Gam}
\eeq
and
\beq
M = m + a_\mu \ga^\mu + b_\mu \ga_5 \ga^\mu
   + \half H_\mn \si^\mn
\quad ,
\label{M}
\eeq
Here,
natural units with $\hbar = c = 1$ are used,
and $i D_\mu \equiv i \partial_\mu - q A_\mu$.
The two terms involving the effective coupling constants
$a_\mu$ and $b_\mu$ violate CPT,
while the three terms involving
$H_{\mu \nu}$, $c_{\mu \nu}$, and $d_{\mu \nu}$
preserve CPT.
All five terms break Lorentz symmetry.\cite{note1}

The recent atomic experiments that test CPT and Lorentz symmetry
express the bounds they obtain in terms of the parameters
$a_\mu$, $b_\mu$, $c_{\mu \nu}$, $d_{\mu \nu}$, and $H_{\mu \nu}$.
This provides a straightforward way of making comparisons across different
types of experiments and avoids problems that can arise
when different physical quantities
(g factors, charge-to-mass ratios, masses, frequencies, etc.)
are used in different experiments.
It is important to keep in mind as well
that each different particle sector in the QED extension
has a set of Lorentz-violating parameters that are independent.
The parameters of the different sectors are distinguished
using superscript labels.
A thorough investigation of possible CPT and Lorentz violation
must look at as many different particle sectors as possible.
The atomic experiments discussed here have obtained bounds on the
parameters for the
electron,\cite{dehmelt99,mittleman99,bkr9798,bk00,eotwash}
muon,\cite{muonium99,muong01,bkl00}
proton,\cite{Hmaser,bkr99}
and
neutron.\cite{dualmaser}
In addition to these,
there are other experiments that provide bounds on
some of the remaining particle sectors,
e.g., neutral mesons\cite{ckpv,mesons} and photons.\cite{ck,photons}

\section{Atomic Experiments}

Before examining the different atomic experiments
that have been performed in recent years,
it is useful to discuss some of the more general
results that have emerged from these investigations.
First, it has become apparent that the
sharp distinction between what are considered Lorentz tests
and CPT tests has been greatly diminished.
Experiments traditionally viewed as Lorentz tests are
also sensitive to CPT and vice versa.
In particular,
it has been shown that it is possible to test CPT in experiments with
particles alone,
which has opened up a whole new arena of CPT tests.
A second general feature of these experiments is the observation that
their sensitivity to CPT and Lorentz violation stems primarily from their
ability to detect very small anomalous energy shifts.
While many of the experiments were originally designed to
measure specific quantities,
such as differences in g factors or
charge-to-mass ratios of particles and antiparticles,
it is now seen that they are most effective as
CPT and Lorentz tests when all of the energy levels in the system
are investigated for possible anomalous shifts.
Indeed,
several new signatures of CPT and Lorentz violation have been
investigated in recent years that were previously overlooked.
Examples of this are given in the following sections.
It has also become common practice to use the relative size of these
anomalous energy shift as figures of merit.
These quantities can in turn be computed in terms of the
parameters in the standard-model extension,
and bounds can be expressed in terms of either.
Finally,
one last common feature these experiments share is that they
all have sensitivity to the Planck scale.

\subsection{Penning-Trap Experiments}

The original experiments with Penning traps were designed to
make high-precision comparisons of the $g$ factors and charge-to-mass ratios of
particles and antiparticles confined within the trap.\cite{penningtests}
These quantities were obtained through measurements of the
anomaly frequency $\om_a$ and the cyclotron frequency $\om_c$.
For example,
$g-2=2\om_a/\om_c$.
The frequencies were measured to $\sim 10^{-9}$ for the electron thereby
determining $g$ to $\sim 10^{-12}$.
In computing these ratios it was not necessary to keep
track of the times when $\om_a$ and $\om_c$ were measured.
It has since been found,
however,
that there are additional signals of possible CPT and Lorentz violation
in this system,
which has led to two new tests being performed.

The first was a reanalysis performed by Dehmelt's group of existing
data for electrons and positrons in a Penning trap.\cite{dehmelt99}
The idea was to look for an instantaneous difference in the
anomaly frequencies of electrons and positrons,
which can be nonzero when CPT and Lorentz symmetry are broken.
(In contrast the instantaneous cyclotron frequencies remain equal
at leading order in the CPT and Lorentz-violation corrections).
Dehmelt's original measurements of $g-2$
did not involve looking for possible instantaneous variations in $\om_a$.
Instead,
the ratio $\om_a/\om_c$ was computed using averaged values.
The new analysis is particularly relevant because it can be shown that
the CPT-violating corrections to the anomaly frequency
$\om_a$ can occur even though the g factor remains unchanged.
The new bound found by Dehmelt's group based on a possible
instantaneous difference in the electron and positron anomaly
frequencies can be expressed in terms of the parameter $b^e_3$,
which is the component of $b^e_\mu$ along the quantization
axis in the laboratory frame.
The bound they obtained is $|b^e_3| \lsim 3 \times 10^{-25}$ GeV.

A second new signal for CPT and Lorentz violation in the electron
sector has been obtained using only data for the electron.\cite{mittleman99}
Here,
the idea is that the CPT and Lorentz-violating interactions depend on
the orientation of the quantization axis in the laboratory frame,
which changes as the Earth turns on its axis.
As a result,
both the cyclotron and anomaly frequencies have small corrections which
cause them to exhibit sidereal time variations.
Such a signal can be measured using electrons alone,
eliminating the need for comparison with positrons.
The bounds in this case must be given with respect to a
nonrotating coordinate system such as celestial equatorial coordinates.
The interactions involve a combination of laboratory-frame components
that couple to the spin of the electron.
This combination is denoted as
$\tilde b_3^e  \equiv b_3^e - m d_{30}^e - H_{12}^e$.
When expressed in terms of components $X$, $Y$, $Z$ in the nonrotating
frame,
the bound obtained by Mittleman {\it et al.} is
$|\tilde b_J^e| \lsim 5 \times 10^{-25} {\rm GeV}$ for $J=X,Y$.

\subsection{Clock-Comparison Experiments}

The classic Hughes-Drever experiments
are atomic clock-comparison tests of Lorentz invariance.\cite{cctests}
These experiments look for relative changes between two ``clock''
frequencies as the Earth rotates.
The ``clock'' frequencies are typically atomic hyperfine or Zeeman transitions.
At the time of the last CPT Meeting in 1998,
the best bounds at leading-order for the proton, neutron and
electron all came from the experiment of Berglund {\it et al.}.
These were, respectively,
$\tilde b_J^p \simeq 10^{-27}$ GeV,
$\tilde b_J^n \simeq 10^{-30}$ GeV,
and $\tilde b_J^e \simeq 10^{-27}$ GeV for $J=X,Y$.
Note that these limits involve bounds on CPT violation in addition
to Lorentz violation.

In the three years since the last meeting,
several new clock-comparison tests have been performed
or are in the planning stages.
For example,
Bear {\it et al.} have used a two-species noble-gas maser to
test for CPT and Lorentz violation in the neutron sector.\cite{dualmaser}
They obtained a new bound
$|\tilde b_J^n| \lsim 10^{-31} {\rm GeV}$ for $J=X,Y$.
This is currently the best bound for the neutron sector.
As spectacular as these bounds are,
however,
it should be pointed out that certain assumptions about the nuclear
configurations must be made in obtaining them.
For this reason,
these bounds should be viewed as good to within about
an order of magnitude.
To obtain cleaner bounds it is necessary to consider
simpler atoms or to perform more sophisticated nuclear modeling.

\subsection{Hydrogen-Antihydrogen Experiments}

The simplest atom one can consider is hydrogen.
Two experiments are being planned at CERN which will
make high-precision spectroscopic measurements of the 1S-2S
transitions in hydrogen and antihydrogen.
These are forbidden transitions with a relative linewidth
of approximately $10^{-15}$.
The idea is ultimately to measure the line center of this
transition to a part in $10^3$ yielding a frequency comparison
between hydrogen and antihydrogen at a level of $10^{-18}$.
An analysis of the 1S-2S transition in the context of the
standard-model extension reveals that the magnetic field plays an important
role
in the sensitivity of the transition to Lorentz and CPT breaking.
For example,
in free hydrogen in the absence of a magnetic field,
the 1S and 2S levels shift by the same amount at leading order.
As a result of this,
there are no leading-order corrections to the 1S-2S transition
frequency in free H or $\bar {\rm H}$.
However,
in a magnetic trap
there are fields that mix the spin states in the
four hyperfine levels.
Since the Lorentz-violating couplings are spin-dependent,
there will be leading-order sensitivity
to Lorentz and CPT violation in comparisons of 1S-2S transitions in
trapped hydrogen and antihydrogen.
However,
these transitions are also field-dependent,
which makes the experimental challenges all the greater.

As an alternative to 1S-2S measurements,
a recent experiment of Phillips {\it et al.} has
considered measurements of the ground-state Zeeman
hyperfine transitions in hydrogen alone.\cite{Hmaser}
It has been shown that these transitions in a hydrogen maser
are sensitive to leading-order Lorentz-violating effects.
Measurements of these transitions have now been made using a double-resonance
technique.\cite{Hmaser}
They give rise to new bounds for the electron and proton.
The bound for the proton alone is $|\tilde b_J^p| \lsim 10^{-27}$ GeV.
Due to the simplicity of hydrogen,
this is an extremely clean bound,
and it is currently the most stringent test
of Lorentz and CPT symmetry for the proton.

\subsection{Spin-Polarized Matter}

A recent experiment at the University of Washington used a spin-polarized
torsion pendulum
\cite{eotwash}
to achieve very high sensitivity to
Lorentz violation in the electron sector.
Its sensitivity comes from the combined effect of a large number
of aligned electron spins.
The experiment uses stacked toroidal magnets with a net
electron spin $S \simeq 8 \times 10^{22}$,
but which have a negligible magnetic field.
The apparatus is suspended on a turntable and a time-varying
harmonic signal is sought.
An analysis of this system shows that in addition to a signal with the
period of the rotating turntable,
the effects of Lorentz and CPT violation would induce additional
time variations with a sidereal period caused by Earth's rotation.
The University of Washington group has analyzed their data
and have obtained a bound on the electron parameters
equal to $|\tilde b_J^e| \lsim 10^{-29}$ GeV for $J=X,Y$ and
$|\tilde b_Z^e| \lsim 10^{-28}$ GeV.\cite{eotwash}
These are now the best Lorentz and CPT bounds for the electron.

\subsection{Muon Experiments}

Experiments with muons involve second-generation leptons and
provide tests of CPT and Lorentz symmetry that are independent
of the tests involving electrons.
There are several different types of experiments with muons
that are currently being conducted,
including muonium experiments\cite{muonium99}
and $g-2$ experiments with muons at Brookhaven.\cite{muong01}
In muonium,
experiments measuring the frequencies
of ground-state Zeeman hyperfine transitions
in a strong magnetic field have the greatest sensitivity
to Lorentz and CPT violation.
A recent analysis has searched for sidereal time variations
in these transitions.
A bound at the level of $| \tilde b^\mu_J| \le 5 \times 10^{-22}$ GeV
has been obtained.\cite{hughes01}
In relativistic $g-2$ experiments using positive muons
with ``magic'' boost parameter $\de = 29.3$,
bounds on Lorentz-violation parameters are possible at
a level of $10^{-25}$ GeV.
These experiments are currently underway at Brookhaven and their results
should be forthcoming in the near future.\cite{deile}

\section{Conclusions}

In summary,
the three years since the first CPT meeting have been a busy time
for the atomic experimentallists conducting tests of CPT and Lorentz symmetry.
Five new sets of bounds have emerged for the electron, proton, neutron, and
muon.
The leading-order bounds from these tests are summarized in Table 1.
All of these bounds are within the range of sensitivity associated
with suppression factors arising from the Planck scale.
However,
as sharp as these bounds are,
there continues to be room for improvement.
Several of the other talks at this meeting describe efforts
to improve these bounds by several orders of magnitude.
In addition,
it should be possible to obtain bounds on many of the parameters
that do not appear in Table 1,
including in particular $Z$ components and timelike components of
the Lorentz-violation parameters.
One promising approach is to conduct clock-comparison tests
in a space satellite.\cite{russell}
For these reasons,
the next few years are likely to be as busy as the previous three.
Atomic experiments will continue to provide
increasingly sharp new tests of CPT and Lorentz
symmetry in matter.

\begin{table}[t]
\begin{center}
\noindent
\renewcommand{\arraystretch}{1.2}
\begin{tabular}{|c|c|c|c|}
\hline\hline
Expt & Sector & Params ($J=X,Y)$ & Bound (GeV)
\\
\hline\hline
Penning Trap & electron & $\tilde b_J^e$ & $5 \times 10^{-25}$ \\[2mm]
\cline{1-4}
Hg-Cs clock  & electron & $\tilde b_J^e$ & $\sim 10^{-27}$ \\[2mm]
\cline{2-4}
comparison & proton & $\tilde b_J^p$ & $\sim 10^{-27}$ \\[2mm]
\cline{2-4}
 & neutron & $\tilde b_J^n$ & $\sim 10^{-30}$ \\[2mm]
\cline{1-4}
He-Xe dual maser & neutron & $\tilde b_J^n$ & $\sim 10^{-31}$ \\[2mm]
\cline{1-4}
H maser & electron & $\tilde b_J^e$ & $10^{-27}$ \\[2mm]
\cline{2-4}
 & proton & $\tilde b_J^p$ & $10^{-27}$ \\[2mm]
\cline{1-4}
Spin Pendulum & electron & $\tilde b_J^e$ & $10^{-29}$ \\[2mm]
&& $\tilde b_Z^e$ & $10^{-28}$ \\[2mm]
\cline{1-4}
Muonium & muon & $\tilde b_J^\mu$ & $2 \times 10^{-23}$ \\[2mm]
\cline{1-4}
Muon g-2 & muon & $\mathaccent 20 b_J^\mu$ & $5 \times 10^{-25}$ \\[2mm]
&&& {\rm (estimated)} \\[2mm]
\hline
\hline
\end{tabular}
\renewcommand{\arraystretch}{1.0}
\caption{Summary of leading-order bounds.}
\vspace{0.2cm}
\end{center}
\end{table}

\section*{Acknowledgments}
This work is supported in part by the National
Science Foundation under grants number PHY-9801869
and PHY-0097982.


\begin{thebibliography}{99}
\bibitem{note1}
Additional terms invloving parameters $e_\mu$, $f_\mu$,
and $g_{\la\mu\nu}$ are ignored.
Gauge invariance excludes these terms in the
standard-model extension.
However,
they could be generated effectively due to strong binding
in a nucleus and are included in a mre general discussion.\cite{kl99}
\end{thebibliography}
\end{document}